\documentclass[aps,print,a4paper,showpacs,12pt,onecolumn,showkeys]{revtex4}
\usepackage{amsfonts}
\usepackage{amsmath}
\usepackage{amssymb}
\usepackage{graphicx}

\input{tcilatex}
\begin{document}

\title{Hirota method for the nonlinear Schr\H{o}dinger equation with an
arbitrary linear time-dependent potential }
\author{Zai-Dong Li$^{1}$, Qiu-Yan Li$^{1}$, Xing-Hua Hu$^{2}$, Zhong-Xi
Zheng$^{3}$, and Yu-Bao Sun$^{1}$}
\affiliation{$^{1}$Department of Applied Physics, Hebei University of Technology, Tianjin
300130, China}
\affiliation{$^{2}$Beijing National Laboratory for Condensed Matter Physics, Institute of
Physics, Chinese Academy of Sciences, Beijing 100080, China }
\affiliation{$^{3}$Department of Physics, North University of China, Taiyuan 030051,
China }

\begin{abstract}
In this paper, a Hirota method is developed for applying to the nonlinear
Schr\"{o}dinger equation with arbitrary time-dependent linear potential
which denotes the dynamics of soliton solutions in quasi-one-dimensional
Bose-Einstein condensation. The nonlinear Schr\"{o}dinger equation is
decoupled to two equations carefully. With a reasonable assumption the one-
and two-soliton solutions are constructed analytically in the presence of an
arbitrary time-dependent linear potential.
\end{abstract}

\maketitle

\section{Introduction}

The realization of Bose-Einstein condensation (BEC) \cite{Anderson,Bradley}
which strongly stimulates the exploration of nonlinear properties of matter
waves have opened a new field of nonlinear atom optics, such as four wave
mixing in BEC's \cite{Deng}, the study of various types of excitations. One
of particular interest is macroscopically excited Bose-Einstein condensed
states, such as vortices \cite{Fetter,Donnely,Matthews} and solitons \cite%
{Kivshar,Ruprecht,Zhang,Muryshev,Denschlag,Bronski,Carr}. The existence of
solitonic solutions is a general feature of nonlinear wave equations. For
the case of an atomic Bose-Einstein condensate, the macroscopic wave
function of the condensate obeys the so-called Gross-Pitaevskii (G-P)
equation, whose nonlinearity result from the interatomic interactions. It is
well known that the G-P equation has of either dark or bright solitons
depending on the repulsive or attractive nature of the interatomic
interactions, respectively. A dark soliton \cite{Dum,Burger} in BEC is a
macroscopic excitation of the condensate which is characterized by a local
density minimum and a phase gradient of the wave function at the position of
the minimum. A bright soliton \cite{Kevin,Khaykovich,Khawaja} in BEC is
expected for the balance between the dispersion and the attractive
mean-field energy. Several methods have been applied to obtain the soliton
solutions of G-P equation with different potential \cite%
{Zhang,Muryshev,Denschlag,Bronski,Carr,Shun,Lizd}, as well as the dynamics
of the excitation of the condensate was discussed. When the longitudinal
dimension of the BEC is much longer than its transverse dimensions which is
the order of its healing length, the G-P equation can be reduced to the
quasi-one-dimensional (quasi-1D) regime. This trapped quasi-low-dimensional 
\cite{Gorlitz} condensates has offered an useful tool for investigating the
nonlinear excitations such as solitons and vortices, which are more stable
than in 3D, where the solitons suffer from the transverse instability and
the vortices can bend. Thus the study of both theory and experiment is very
important for the soliton excitations in quasi-low-dimensional BECs.

In this paper, we consider the mean-field model of a quasi-1D BEC trapped in
a linear time-dependent potential which is given by 
\begin{equation}
i\hbar\frac{\partial}{\partial T}\Psi=-\frac{\hbar^{2}}{2m}\frac{\partial^{2}%
}{\partial X^{2}}\Psi+Xf(T)\Psi+g\left\vert \Psi\right\vert ^{2}\Psi,
\label{NLS}
\end{equation}
where $\dint \left\vert \Psi\right\vert ^{2}dX=N$ is the number of atoms in
the condensate, $f(T)$ is the arbitrary function of time $T$, the
interacting constant of two-atom is given by $g$ $=2\hbar^{2}a/ml_{0}^{2}$ (%
\cite{Petrov}) with $m$ the mass of the atom, $a$ the $s$-wave scattering
length ($a>0$ for repulsive interaction; while $a<0$ for attractive
interaction), and $l_{0}\equiv \sqrt{\hbar/m\omega_{0}}$ denoting the
characteristic length extension of the ground state wave function of
harmonic oscillator. Making a dimensionless transformation, we can rewrite
equation (1) as%
\begin{equation}
i\frac{\partial}{\partial t}\psi+\frac{1}{2}\frac{\partial^{2}}{\partial
x^{2}}\psi+xf\left( t\right) \psi+\mu\left\vert \psi\right\vert ^{2}\psi=0,
\label{NLS1}
\end{equation}
where $x$ is measured in units of $l_{0}$, $t$ in units of $ml_{0}^{2}/\hbar$%
, $\psi$ in units of the square root of $Nl_{0}$, the interaction constant $%
\mu$ is defined as $\mu=-2Nl_{0}a$, and $f\left( t\right) =-\frac{ml_{0}^{3}%
}{\hbar^{2}}f(\frac{t}{\hbar/ml_{0}^{2}})$. The exact soliton solutions of
Eq. (\ref{NLS1}) can been constructed by the inverse scattering method \cite%
{Chen} and F-expansion method \cite{ZhangJF}. It should be noted that in the
absence of the linear potential, i.e. $xf\left( t\right) =0$, a Hirota
method can be applied to Eq. (\ref{NLS1}) directly for getting the bright
and dark soliton solutions. However, with the consideration of the linear
potential the Hirota method should be developed carefully. This is our
purpose in the present paper. With a reasonable assumption we demonstrate
how to construct the exact one- and two-soliton solutions of Eq. (\ref{NLS1}%
) in terms of this developed technique.

\section{One-soliton solution}

Now we introduce the main idea of the Hirota method briefly. Firstly, it
apply a direct transformation to the nonlinear equation. Then in terms of
the reasonable assumption the nonlinear equation can be decoupled to two
equations from which the one- and two-soliton solutions can be constructed
effectively. To this purpose we consider the following transformation

\begin{equation}
\psi=\frac{G(x,t)}{F(x,t)},  \label{hirota1}
\end{equation}
where $G\left( x,t\right) $ is complex function and $F\left( x,t\right) $ is
a real function. With this transformation Eq. (\ref{NLS1}) becomes%
\begin{equation}
F\left( iD_{t}+\frac{1}{2}D_{x}^{2}\right) G\cdot F+G\cdot F^{2}\cdot
xf(t)-G(\frac{1}{2}D_{x}^{2}F\cdot F-\mu\overline{G}G)=0,  \label{hirota2}
\end{equation}
where the overbar denotes the complex conjugate, $D_{t}$ and $D_{x}^{2}$ are
called the Hirota bilinear operators defined as 
\begin{equation}
D_{x}^{m}D_{t}^{n}G\left( x,t\right) \cdot F\left( x,t\right) =\left( \frac{%
\partial}{\partial x}-\frac{\partial}{\partial x^{\prime}}\right) ^{m}\left( 
\frac{\partial}{\partial t}-\frac{\partial}{\partial t^{\prime}}\right)
^{n}G\left( x,t\right) F\left( x^{\prime},t^{\prime}\right) \left.
{}\right\vert _{x=x^{\prime},t=t^{\prime}}.  \label{oper1}
\end{equation}
If the term $xf(t)=0$ Eq. (\ref{hirota2}) reduces to the normal nonlinear
equation which can be decoupled easily to two equations. In the presence of
the term $xf(t)$ we should deal with Eq. (\ref{hirota2}) carefully. Many
attempts show that Equation (\ref{hirota2}) can be decoupled as 
\begin{align}
\frac{1}{2}D_{x}^{2}F\cdot F-\mu\overline{G}G & =0,  \notag \\
\left( iD_{t}+\frac{1}{2}D_{x}^{2}+xf(t)\right) G\cdot F & =0,
\label{hirota3}
\end{align}
in which the spatial and time dependence term $xf(t)$ will give a difficulty
for getting solutions as shown below. Now the Eq. (\ref{hirota3}) has made
the Eq. (\ref{NLS1}) to the normal procedure of Hirota method for getting
the exact soliton solutions. By making a series of suitable assumption for
the expression of $G$ and $F$, the exact one- and two-soliton solution can
be obtained analytically. In order to obtain the bright one-soliton solution
of Eq. (\ref{NLS1}) which correspond to the case $a<0$, we proceed in the
standard assumption 
\begin{equation}
G=\chi G_{1},\text{ }F=1+\chi^{2}F_{1},  \label{ansaz1}
\end{equation}
where $\chi$ is an arbitrary parameter which will be absorbed in expressing
the soliton solution in the following sections. Substituting Eq. (\ref%
{ansaz1}) into Eq. (\ref{hirota3}), then collecting the coefficients with
same power in $\chi$, we have

(1) for the coefficient of $\chi$

\begin{equation}
\left( iD_{t}+\frac{1}{2}D_{x}^{2}+xf(t)\right) G_{1}\cdot1=0,  \label{term1}
\end{equation}

(2) for the coefficient of $\chi^{2}$%
\begin{equation}
D_{x}^{2}F_{1}-\mu\overline{G}_{1}G_{1}=0,  \label{term2}
\end{equation}

(3) for the coefficient of $\chi^{3}$%
\begin{equation}
\left( iD_{t}+\frac{1}{2}D_{x}^{2}+xf(t)\right) G_{1}\cdot F_{1}=0,
\label{term3}
\end{equation}

(4) for the coefficient of $\chi^{4}$%
\begin{equation}
D_{x}^{2}F_{1}\cdot F_{1}=0.  \label{term4}
\end{equation}
Using the definition (\ref{oper1}) the above equations can be expressed in
detail. For example, in order to satisfy Eq. (\ref{term1})\ we can assume $%
G_{1}$ has the form

\begin{equation}
G_{1}=e^{\eta_{1}}.  \label{solution1}
\end{equation}
Substituting Eq. (\ref{solution1}) into Eq. (\ref{term1}) we have%
\begin{equation}
i\eta_{1t}+\frac{1}{2}\eta_{1xx}+\frac{1}{2}\eta_{1x}^{2}+xf(t)=0.
\label{para}
\end{equation}
Because of the presence of the term $xf(t)$ there are some difficulties for
getting general solutions of the above equation. For some conveniences in
this paper we assume $\eta_{1}$ has the form 
\begin{equation}
\eta_{1}=P_{1}(t)x+\Omega_{1}(t),  \label{para1}
\end{equation}
with the time-dependent functions $P_{1}(t)$ and $\Omega_{1}(t)$ to be
determined. With the restriction, i.e., Eqs. (\ref{term2}) and (\ref{term4})
we get 
\begin{equation}
F_{1}=\exp(\eta_{1}+\overline{\eta}_{1}+A_{11}),  \label{solution2}
\end{equation}
where%
\begin{equation*}
A_{11}=\ln\frac{\mu}{\left( P_{1}+\overline{P}_{1}\right) ^{2}}.
\end{equation*}
Substituting the solutions (\ref{solution1}) and (\ref{solution2}) into Eq. (%
\ref{term3}) we obtain the equations of $P_{1}(t)$ and $\Omega_{1}(t)$ as 
\begin{equation}
\left[ iP_{1,t}(t)+f(t)\right] x+i\Omega_{1,t}+\frac{1}{2}P_{1}^{2}=0,
\label{condition1}
\end{equation}
form which we can determine the expression of $P_{1}(t)$ and $\Omega_{1}(t)$%
. It is obvious that Eq. (\ref{condition1}) implies the natural conditions 
\begin{align}
iP_{1,t}(t)+f(t) & =0,  \notag \\
i\Omega_{1,t}(t)+\frac{1}{2}P^{2}(t) & =0.  \label{para2}
\end{align}
Solving the above two equations, we get the expression of $P_{1}$ and $%
\Omega_{1}$ as 
\begin{align}
P_{1} & =i\int_{0}^{t}f(\tau)d\tau+\xi_{10},  \notag \\
\Omega_{1} & =\frac{i}{2}\int_{0}^{t}P_{1}^{2}\left( \tau\right)
d\tau+\zeta_{10},  \label{para3}
\end{align}
where $\xi_{10}$ and $\zeta_{10}$ are complex parameters in general. With
the Eqs. (\ref{solution1}), (\ref{solution2}) and (\ref{hirota1}), after
absorbing $\chi$, the bright one-soliton solution of Eq. (\ref{NLS1}) can be
derived as 
\begin{equation}
\psi=\frac{e^{\eta_{1}}}{1+e^{\eta_{1}+\overline{\eta}_{1}+A_{11}}},
\label{onesoliton}
\end{equation}
where%
\begin{align}
\eta_{1} & =\left[ i\int_{0}^{t}f(\tau)d\tau+\xi_{10}\right] x+\frac{i}{2}%
\int_{0}^{t}P_{1}^{2}\left( \tau\right) d\tau+\zeta_{10}.  \notag \\
A_{11} & =\ln\frac{\mu}{\left( \xi_{10}+\overline{\xi}_{10}\right) ^{2}}.
\label{onepapra1}
\end{align}
In the case of $f(t)=0$, one-soliton solution (\ref{onesoliton}) can reduce
to the solutions of the normal nonlinear Schr\H{o}dinger equation. When $%
f(t)=constant$ the solution (\ref{onesoliton}) is the same results (\cite%
{Chen}) reported earlier. As $f(t)=b_{1}+l\cos\left( \omega t\right) $, the
solution (\ref{twosoliton}) is the same results (\cite{ZhangJF}). From Eq. (%
\ref{onepapra1}) we can see the linear time-dependent potential can change
the soliton velocity and frequency.

\section{Two-soliton solution}

In this section we will give the analytical expression of two soliton
solution of Eq. (\ref{NLS1}). To this purpose we now assume that 
\begin{equation}
G=\chi G_{1}+\chi^{3}G_{2},\text{ }F=1+\chi^{2}F_{1}+\chi^{4}F_{2}.
\label{ansaz2}
\end{equation}
By employing the same procedure before we obtain the following set of
equations from Eq. (\ref{hirota3}), corresponding to the different powers of 
$\chi$

(1) for the coefficient of $\chi$%
\begin{equation}
\left( iD_{t}+\frac{1}{2}D_{x}^{2}+xf(t)\right) G_{1}\cdot1=0,  \label{th1}
\end{equation}

(2) for the coefficient of $\chi^{2}$%
\begin{equation}
D_{x}^{2}F_{1}\cdot1=\mu\overline{G}_{1}G_{1},  \label{th2}
\end{equation}

(3) for the coefficient of $\chi^{3}$%
\begin{equation}
\left( iD_{t}+\frac{1}{2}D_{x}^{2}\right) (G_{1}\cdot
F_{1}+G_{2}\cdot1)+xf(t)\left( G_{1}F_{1}+G_{2}\right) =0,  \label{th3}
\end{equation}

(4) for the coefficient of $\chi^{4}$%
\begin{equation}
D_{x}^{2}F_{1}\cdot F_{1}+2D_{x}^{2}F_{2}\cdot1=2\mu(\overline{G}_{1}G_{2}+%
\overline{G}_{2}G_{1}),  \label{th4}
\end{equation}

(5) for the coefficient of $\chi^{5}$%
\begin{equation}
\left( iD_{t}+\frac{1}{2}D_{x}^{2}\right) (G_{1}\cdot F_{2}+G_{2}\cdot
F_{1})+xf(t)\left( G_{1}F_{2}+G_{2}F_{1}\right) =0,  \label{th5}
\end{equation}

(6) for the coefficient of $\chi^{6}$%
\begin{equation}
D_{x}^{2}F_{1}\cdot F_{2}=\mu\overline{G}_{2}G_{2},  \label{th6}
\end{equation}

(7) for the coefficient of $\chi^{7}$%
\begin{equation}
\left( iD_{t}+\frac{1}{2}D_{x}^{2}+xf(t)\right) G_{2}\cdot F_{2}=0,
\label{th7}
\end{equation}

(8) for the coefficient of $\chi^{8}$%
\begin{equation}
D_{x}^{2}F_{2}\cdot F_{2}=0,  \label{th8}
\end{equation}
As discussed in the one-soliton solution, we can solve the equations from (%
\ref{th1}) to (\ref{th8}) in turn for getting the expression of $G$ and $F$.
In order to construct the two-soliton solution of (\ref{NLS1}) we assume $%
G_{1}$ has the form

\begin{equation}
G_{1}=\exp\eta_{1}+\exp\eta_{2},  \label{N2solution1}
\end{equation}
where%
\begin{equation*}
\eta_{j}=P_{j}(t)x+\Omega_{j}(t),
\end{equation*}
in which the time-dependent functions $P_{j}(t)$ and $\Omega_{j}(t)$, $%
j=1,2, $ to be determined. Substituting Eq. (\ref{N2solution2}) into the
relation (\ref{th1}) we have%
\begin{equation*}
\left\{ \left[ iP_{1,t}+f(t)\right] x+i\Omega_{1,t}+\frac{1}{2}%
P_{1}^{2}\right\} \exp\eta_{1}+\left\{ \left[ iP_{2,t}+f(t)\right]
x+i\Omega_{2,t}+\frac{1}{2}P_{2}^{2}\right\} \exp\eta_{2}=0,
\end{equation*}
which implies that%
\begin{align*}
iP_{j,t}+f(t) & =0, \\
i\Omega_{j,t}+\frac{1}{2}P_{j}^{2} & =0.
\end{align*}
where $j=1,2$. From the above equations one can find the solutions%
\begin{align*}
P_{j}\left( t\right) & =i\int_{0}^{t}f(\tau)d\tau+\xi_{j0}, \\
\text{ }\Omega_{j}(t) & =\frac{i}{2}\int_{0}^{t}P_{j}^{2}\left( \tau\right)
d\tau+\zeta_{j0},
\end{align*}
where $\xi_{j0}$ and $\zeta_{j0}$, $j=1,2$, are complex parameters in
general. Combining Eq. (\ref{N2solution1}) with Eq. (\ref{th2}) we obtain
the expression of $F_{1}$ as%
\begin{align}
F_{1} & =\exp(\eta_{1}+\overline{\eta}_{1}+A_{11})+\exp(\eta_{2}+\overline{%
\eta}_{1}+A_{21})  \notag \\
& +\exp(\eta_{1}+\overline{\eta}_{2}+A_{12})+\exp(\eta_{2}+\overline{\eta }%
_{2}+A_{22}),  \label{N2solution2}
\end{align}
\qquad where%
\begin{equation*}
A_{mn}=\ln\frac{\mu}{\left( P_{m}+\overline{P}_{n}\right) ^{2}},\text{ }%
m,n=1,2.
\end{equation*}
With the help of Eqs. (\ref{N2solution1}) and (\ref{N2solution2}) we can
simplify Eq. (\ref{th3}) as%
\begin{align}
& iG_{2t}+\frac{1}{2}G_{2xx}+xf(t)G_{2}  \notag \\
& =\frac{\mu\left( P_{2}-P_{1}\right) ^{2}}{\left( P_{1}+\overline{P}%
_{1}\right) \left( P_{2}+\overline{P}_{1}\right) }e^{\eta_{1}+\bar{\eta }%
_{1}+\eta_{2}}+\frac{\mu\left( P_{2}-P_{1}\right) ^{2}}{\left( P_{1}+%
\overline{P}_{2}\right) \left( P_{2}+\overline{P}_{2}\right) }e^{\eta_{1}+%
\bar{\eta}_{2}+\eta_{2}},  \label{g2a}
\end{align}
which shows that the expression of $G_{2}$ has the form 
\begin{equation}
G_{2}=\exp(\eta_{1}+\overline{\eta}_{1}+\eta_{2}+\delta_{1})+\exp(\eta _{1}+%
\overline{\eta}_{2}+\eta_{2}+\delta_{2}),  \label{N2solution3}
\end{equation}
where the parameter $\delta_{j}$, $j=1,2,$ is given by 
\begin{equation}
\delta_{j}=\ln\left( \frac{B_{j}}{\gamma_{j}}\right) ,j=1,2,
\label{twopara1}
\end{equation}
\begin{align*}
\gamma_{1} & =\left( P_{1}+\overline{P}_{1}\right) \left( P_{2}+\overline{P}%
_{1}\right) ,\gamma_{2}=\left( P_{1}+\overline{P}_{2}\right) \left( P_{2}+%
\overline{P}_{2}\right) , \\
B_{1} & =\frac{\mu\left( P_{2}-P_{1}\right) ^{2}}{\left( P_{1}+\overline{P}%
_{1}\right) \left( P_{2}+\overline{P}_{1}\right) },\text{ }B_{2}=\frac{%
\mu\left( P_{2}-P_{1}\right) ^{2}}{\left( P_{1}+\overline {P}_{2}\right)
\left( P_{2}+\overline{P}_{2}\right) }.
\end{align*}
Now we have obtained the expression of $G_{1}$, $G_{2}$, and $F_{1}$ in Eq. (%
\ref{ansaz2}). Substituting Eqs. (\ref{N2solution1}), (\ref{N2solution2})
and (\ref{N2solution3}) into Eq. (\ref{th4}) we obtain

\begin{equation}
F_{2}=e^{\eta_{1}+\overline{\eta}_{1}+\eta_{2}+\overline{\eta}_{2}+\kappa},
\label{F2}
\end{equation}
where%
\begin{equation*}
\kappa=\ln\frac{\mu^{2}\left\vert P_{2}-P_{1}\right\vert ^{4}}{\left( P_{1}+%
\overline{P}_{1}\right) ^{2}\left( P_{2}+\overline{P}_{2}\right)
^{2}\left\vert P_{1}+\overline{P}_{2}\right\vert ^{4}}.
\end{equation*}
With the help of Eqs. (\ref{N2solution1}), (\ref{N2solution2}), (\ref%
{N2solution3}), and (\ref{F2}) one can find the Eqs. (\ref{th5}) and (\ref%
{th8}) are satisfied to the moment after a tedious calculation. So the
two-soliton solutions has the form%
\begin{equation}
\psi_{2}=\frac{G}{F},  \label{twosoliton}
\end{equation}
where%
\begin{align*}
G & =e^{\eta_{1}}+e^{\eta_{2}}+e^{\eta_{1}+\overline{\eta}_{1}+\eta
_{2}+\delta_{1}}+e^{\eta_{1}+\overline{\eta}_{2}+\eta_{2}+\delta_{2}}, \\
F & =1+e^{\eta_{1}+\overline{\eta}_{1}+A_{11}}+e^{\eta_{2}+\overline{\eta }%
_{1}+A_{21}}+e^{\eta_{1}+\overline{\eta}_{2}+A_{12}}+e^{\eta_{2}+\overline{%
\eta}_{2}+A_{22}}+e^{\eta_{1}+\overline{\eta}_{1}+\eta _{2}+\overline{\eta}%
_{2}+\kappa}.
\end{align*}%
\begin{align*}
P_{j}\left( t\right) & =i\int_{0}^{t}f(\tau)d\tau+\xi_{j0}, \\
\text{ }\Omega_{j}(t) & =\frac{i}{2}\int_{0}^{t}P_{j}^{2}\left( \tau\right)
d\tau+\zeta_{j0}, \\
\delta_{j} & =\ln\left( \frac{B_{j}}{\gamma_{j}}\right) ,
\end{align*}
where $j=1,2$, and 
\begin{align*}
\gamma_{1} & =\left( \xi_{10}+\overline{\xi}_{10}\right) \left( \xi _{20}+%
\overline{\xi}_{10}\right) , \\
\gamma_{2} & =\left( \xi_{10}+\overline{\xi}_{20}\right) \left( \xi _{20}+%
\overline{\xi}_{20}\right) , \\
B_{1} & =\frac{\mu\left( \xi_{20}-\xi_{10}\right) ^{2}}{\left( \xi _{10}+%
\overline{\xi}_{10}\right) \left( \xi_{20}+\overline{\xi}_{10}\right) },%
\text{ } \\
B_{2} & =\frac{\mu\left( \xi_{20}-\xi_{10}\right) ^{2}}{\left( \xi _{10}+%
\overline{\xi}_{20}\right) \left( \xi_{20}+\overline{\xi}_{20}\right) },
\end{align*}%
\begin{align*}
\kappa & =\ln\frac{\mu^{2}\left\vert \xi_{20}-\xi_{10}\right\vert ^{4}}{%
\left( \xi_{10}+\overline{\xi}_{10}\right) ^{2}\left( \xi_{20}+\overline{\xi}%
_{20}\right) ^{2}\left\vert \xi_{10}+\overline{\xi}_{20}\right\vert ^{4}}, \\
A_{mn} & =\ln\frac{\mu}{\left( \xi_{m0}+\overline{\xi}_{n0}\right) ^{2}},%
\text{ }m,n=1,2.
\end{align*}
When $f(t)=0$, the solution (\ref{twosoliton}) denotes two soliton
interaction of the normal nonlinear Schr\H{o}dinger equation. The new
expression (\ref{twosoliton}) implies that Hirota method has more advantage
for getting new soliton solutions as well.

\section{Conclusion}

In this paper, we investigate the soliton solutions of the nonlinear Schr%
\"{o}dinger equation with an arbitrary time-dependent linear potential which
denotes the dynamics of quasi-one-dimensional Bose-Einstein condensation. A
developed Hirota method is applied carefully to the nonlinear Schr\"{o}%
dinger equation. In terms of this developed technique we decoupled the
nonlinear Schr\"{o}dinger equation into two equations. Moreover, with a
reasonable assumption the exact one- and new two-soliton solutions are
constructed effectively. Our soliton interaction will have useful
application in the studies of Bose-Einstein condensation and optics
communication in which the nonlinear Schr\"{o}dinger equation are used
widely.

\section{Acknowledgement}

This work is supported by the Natural Science Foundation of China No.
10647122, the Natural Science Foundation of Hebei Province of China Grant
No. A2007000006, the Foundation of Education Bureau of Hebei Province of
China No. 2006110, and the key subject construction project of Hebei
Provincial University of China.

\end{document}